# Forecasting Transformative AI: An Expert Survey

Ross Gruetzemacher, David Paradice, Kang Bok Lee
Auburn University, Harbert College of Business, Department of Systems and Technology

**Abstract**

Transformative AI technologies have the potential to reshape critical aspects of society in the near future. In order to properly prepare policy initiatives for the arrival of such technologies accurate forecasts and timelines are necessary. A survey was administered to attendees of three AI conferences in 2018 (ICML, IJCAI and the HLAI conference). The survey included calibration questions, questions for estimating AI capabilities over the next decade, questions for forecasting five scenarios of transformative AI and questions concerning the impact of computational resources in AI research. *Respondents indicated a median of 21.5% of human tasks (i.e., all tasks that humans are currently paid to do) can be feasibly automated now, and that this figure would rise to 40% in 5 years and 60% in 10 years. Median forecasts indicated a 50% probability of AI systems being capable of automating 90% of current human tasks in 25 years and 99% of current human tasks in 50 years*. The conference of attendance was found to have a statistically significant impact on all forecasts, with attendees of HLAI providing more optimistic timelines with less uncertainty. These findings suggest that AI experts expect major advances in AI technology to continue over the next decade.

## 1. Introduction

Forecasting human-level machine intelligence has been a topic of interest since as early as 1972, when Michie polled participants of a lecture series at the University College of London. The majority of respondents indicated that something akin to human-level machine intelligence would take 50 years or greater to realize (Michie 1973). However, it was not until recently that the topic began to gain serious consideration by academic researchers. Eleven more surveys have been administered since 2006 (Grace 2015, Zhang and Dafoe 2019), four of which were ultimately reported in academic outlets (Baum, Goertzel et al. 2011, Müller and Bostrom 2016, Grace, Salvatier et al. 2018, Walsh 2018).

Baum et al. conducted a survey at the 2[nd] Conference on Artificial General Intelligence in 2009 (Baum, Goertzel et al. 2011). This study involved only 21 participants. The majority of these experts expected human-level artificial intelligence (HLAI) to be developed in the coming decades, and the researchers concluded that the possibility of developing HLAI in the relative near-term should be given serious consideration. Despite the low sample size, the study was able to assess participants' opinions further through targeted follow-up questions emailed after the initial survey. 18 of the 21 participants completed these follow-up questions. This qualitative element of the survey enabled the researchers to explore questions regarding the impact of artificial general intelligence (AGI).

In 2012 Müller and Bostrom conducted a survey that consisted of experts from four different samples (Müller and Bostrom 2016). These samples included the top 100 authors in artificial intelligence by number of citations, the members of the Greek Association for Artificial Intelligence, participants of the 2011 Philosophy and Theory of AI conference, participants of the 5th Conference on Artificial General Intelligence and participants of the AGI Impacts 2012 conference. They collected a total of 170 responses, over 40% of which were accounted for by the AGI groups. This study elicited expert opinion in the form of probability distributions, however, only descriptive statistics were used in the analysis. The median response for 50% probability of high-level machine intelligence (HLMI) was 2040, with the AGI groups having the shortest responses. Experts further expect that once HLMI has been achieved, systems will move toward superintelligence[1] within 30 years.

Walsh also conducted a survey of AI experts whose participants were comprised of three groups: 1) authors of the 2015 annual conference for the Association for the Advancement of Artificial Intelligence and authors of the 2011 International Joint Conference on Artificial Intelligence; 2) fellows of the Institute for Electrical and Electronics Engineers and authors of the 2016 International Conference on Robotics and Automation; 3) online readers of an article on an AI system, *Libratus,* that defeated professionals in the card game of poker. The responses were 200, 101 and 548 for each of the groups, respectively. This survey primarily concerned with the risks of occupations being automated. However, it did include one question regarding when computers may reach HLMI. Experts predicted a significant number of occupations to be at risk of automation over the next two decades but were more cautious than non-experts. Probability distributions were elicited for the HLMI question, and results were depicted as cumulative distribution functions visually. Medians for the 50% probability of HLMI were reported for each of the groups with 2065 and 2061 for the first two groups. The last group, i.e. non-experts, gave a median value starkly lower of 2039.

Grace et al. conducted a widely discussed survey in 2016 which surveyed authors from the 2015 Neural Information Processing Systems conference and authors of the 2015 International Conference on Machine Learning. The survey received 352 responses. This study elicited expert opinion as probability distributions and treated the data accordingly in analysis[2]. This study had some interesting results. For one, the study evaluated framing effects by questioning half of the participants for fixed probabilities and half of the participants in fixed years. There was thought to be a difference in forecasts based on framing, however, it was ambiguous as to which framing was more accurate and there was no statistical analysis to support this conclusion. The combined forecasts were reported and the median 50% probability for the development of HLMI was 45 years. In addition to the HLMI question, a question was asked about when AI will outperform humans at all jobs. While worded to be logically equivalent to the HLMI question, the median 50% probability for this question was 120 years.

---

[1] Superintelligence refers to a recursively self-improving HLMI (Bostrom 2014).
[2] This study does the same, and improves slightly on the technique used by Grace et al. in using a multivariate model. This is discussed further in the following pages.

The primary concept of interest in this survey is transformative AI[3]. For the purposes of this study we define transformative AI to be AI that significantly transforms society by replacing humans for a large portion (*i.e.*, 50% or greater) of economically useful work. As such, transformative AI would include notions of AGI, HLMI, human-level machine intelligence, human-level artificial intelligence (HLAI) and machine superintelligence[4], as well as narrow AI technologies that also meet this definition. Furthermore, the concept of transformative AI described here is also consistent with Drexler's comprehensive AI services (CAIS) model (Drexler 2019). Given the definition used here, there is substantial ambiguity regarding transformative AI that poses a challenge in forecasting. In order to counter this challenge, questions are asked regarding a number of scenarios for transformative AI. To this end, we create a distinction between narrow AI and broadly capable AI systems. Narrow AI systems are systems created specifically for a unique task or class of tasks. A broadly capable system refers to a single algorithm that can be used to achieve a wide variety of tasks. Narrow systems can be thought to comprise the transformative AI systems described by CAIS. Broadly capable systems would include transformative AI systems as they approach the previously described notions of AGI.

## 2. Methods

This paper details the results of a survey administered at three AI conferences over the summer of 2018[5]; the International Conference on Machine Learning (ICML), the International Joint Conference on Artificial Intelligence (IJCAI) and the Joint Multi-Conference on Human-Level Artificial Intelligence (HLAI). ICML and IJCAI were held consecutively in Stockholm, Sweden from July 10th-19th. HLAI was held in Prague, Czech Republic from August 21st-25th. HLAI included the annual Artificial General Intelligence Conference, the annual Biologically Inspired Cognitive Architectures Conference and the annual Workshop on Neural-Symbolic Learning and Reasoning.

Attendees of these conferences were selected at random and asked to participate in the survey. Agreeable participants were sent an invite link via email to complete the survey. Participants were sent two reminder emails if they had not yet completed the survey. 406 email addresses were collected; 286 from ICML/IJCAI and 120 from HLAI. Of these, 203 responses were received, 164 of which were completed in full; 119 from ICML/IJCAI and 45 from HLAI. The overall completion rate was 40.4%. The completion rate for ICML/IJCAI was 37.5% while completion rate for the HLAI conference was 50.6%.

**Questions**

The survey[6] was comprised of eight content questions, four calibration questions and four general questions (*i.e.*, first name, last name, employer and job title). There were four calibration questions, each of which asked for a probability that a certain near-term probable AI milestone would occur. The content questions were refined through interviews with machine learning experts. The first six concerned the use of AI to complete tasks that humans are currently paid to do. These questions were split into four pages. Each of these pages included specific instructions, including definitions, for the questions shown on the page. The final two questions concerned the relationship between computational resources and progress in AI research and were shown on the same page. This page included a figure and specific instructions. Comment[7] boxes were made available for all content question pages.

---

[3] There is no consensus yet in the AI strategy community as to what level of societal transformation should be used the threshold for transformative AI.
[4] For the remainder of the document we will refer to all such notions of intelligent systems as notions of AGI.
[5] The survey is part of a broader, mixed-methods study including expert interviews. Work on this is forthcoming.
[6] The survey can be taken at https://www.surveymonkey.com/r/2018_AI_survey.

The four calibration questions were included in order to assess experts' aptitude for forecasting. To do this we asked four questions about the probability of AI milestones that could plausibly occur within the next three years. Two of the questions concerned the availability of level four and level five autonomous vehicles by the end of 2019 and 2021, respectively, a question about the quality of DeepMind's milestones and a final question which was a very near-term question and different for each of the two locations where the survey was solicited. In the first location participants were asked about the probability of OpenAI beating a human professional team in the multiplayer game Dota 2 and in the second location participants were asked about the probability of deep fakes being used to affect the upcoming American elections.

The first six content questions were dependent on the definition of human tasks. We define human tasks as all unique tasks that humans are currently paid to do. We differentiate human tasks from jobs in that jobs are comprised of human tasks. For example, an AI system(s) may not replace a lawyer entirely but may be able to accomplish 50% of the tasks a lawyer typically performs. The first of these six questions asked respondents to estimate the percentage of human tasks feasible for being completed by AI systems at fixed years (*i.e.*, at present, in five years and in ten years). The remaining of these questions asked for forecasts concerning completion of human tasks in number of years at probabilities of 10%, 50% and 90% for five different transformative AI scenarios. This format enabled fitting a probability distribution for each respondent's forecast.

Although we were interested in transformative AI, we did not use the term in the survey. Nor did we use any terms associated with notions of AGI. Rather, we asked for forecasts of AI systems capable of completing a percentage of human tasks at or above the level of a typical human. We asked participants for forecasts of both narrow AI systems and broadly capable AI systems. For forecasts concerning what we consider as narrow AI systems, we simply asked for forecasts of when it would be feasible for AI systems to collectively be able to accomplish the given percentage of human tasks at or above the level of a typical human. We asked for forecasts of narrow AI systems capable of completing 50%, 90% and 99% of human tasks and for forecasts of broadly capable AI systems capable of completing 90% and 99% of such tasks[8]. These were asked of respondents on three pages, with questions being grouped by common percentage and ascending from 50% to 99%.

The final two questions concerned forecasting trends in AI computation and the impact of increased computational resources. In the first of these questions, participants were shown a plot from recent research at OpenAI (Amodei and Hernandez 2018). It is a semi-log plot of computational costs of training for a number of AI milestones over the previous six years. It reveals a doubling in training costs every 3.5 months. Participants were then asked to give a probability of this trend continuing for five years and ten years. The second question asked participants how much they expected AI progress to speed up if researchers were all given access to unlimited computational resources.

---

[7] A thorough discussion of respondents' comments will be included in forthcoming work. This will also explore the interviews conducted concurrently with the survey based on the same questions.
[8] We do not envision a broadly capable AI system as being feasible for only 50% of human tasks.

Müller and Bostrom fixed the maximum forecast at 5,000 years and also offered participants an option of never (Müller and Bostrom 2016). For this survey, we left the questions open-ended, but during analysis we also applied the 5,000-year threshold and classified answers above this threshold as indicative of never. Other inconsistencies included values misordered forecasts and blank entries[9]. The frequency of these cases is reported in the results, but such answers could not be analyzed numerically and were thus discarded for further analysis. Of those who completed the survey, 119 comments were left by 54 participants[10].

## Results

### Fixed Years Human Tasks

The first of the human tasks related questions asked for estimations of the feasibility of automating human tasks at fixed years; first for the present, then for five years and then for ten years. The data from the three parts of this question is presented in Figure 1a and Figure 1b. Participants were asked only for point estimates in these questions. Consequently, only descriptive statistics are reported. Figure 1a depicts histograms for responses on the percentage of human tasks feasible to be automated now, in five years and in ten years. Figure 1b depicts box plots for the same. The median percentage of tasks feasible for automation now is 21.5%. This increases to 40.0% in five years and 60.0% in 10 years.

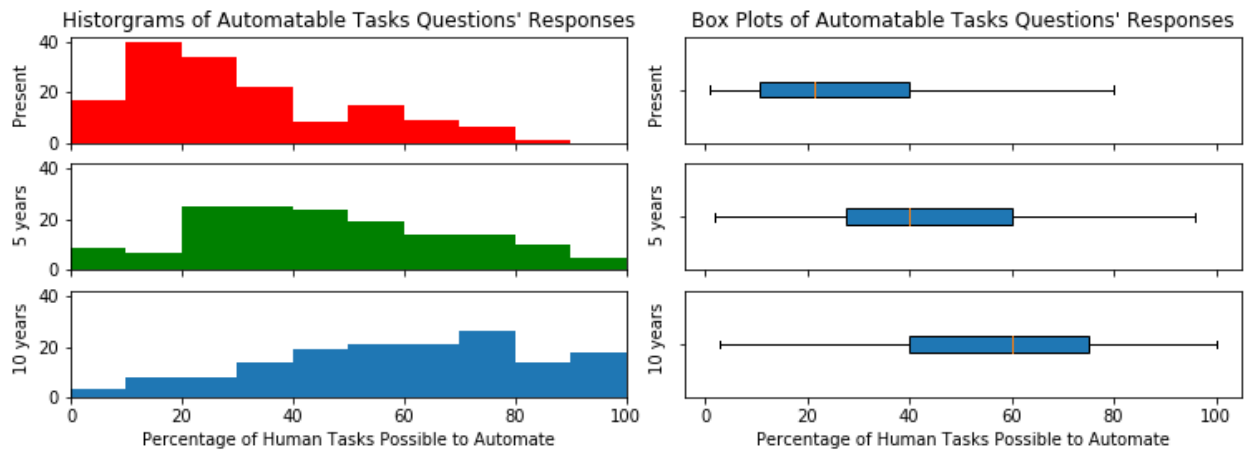

**Figure 1:** left – a) histograms for responses on human tasks feasibly automatable now, in five years and in ten years; right – b) box plots for responses on human tasks feasibly automatable now, in five years and in ten years.

### Forecasts

The next five questions concern forecasts for five different transformative AI scenarios. The median, mean and median aggregate forecast values for 10%, 50% and 90% probabilities can be seen in Table I for each of the five scenarios. Data from individual forecasts was fit to a gamma distribution using least squares (Grace, Salvatier et al. 2018). Numerous techniques exist for aggregation of probability distributions collected via elicitation of expert opinion (Clemen and

---

[9] An answer for each question was required to proceed through the survey. However, some participants exploited the fact that only one answer to each question was sufficient to proceed despite questions requiring multi-part answers.
[10] This information is of interest and will be addressed in forthcoming work – see also footnote 5.

Winkler 1999). Here, we used a median aggregation technique (Hora, Fransen et al. 2013) that has recently been shown to have advantages over the more common quantile aggregation (Vincent 1912). Each of the aggregate forecasts is shown in Figure 2a. Figure 2b depicts a comparison of aggregate forecasts from HLAI attendees and ICML/IJCAI attendees for a 99% broadly capable system.

**Table I: Median, Mean and Median Aggregate Forecast Results**

|  | Probability | System Type | | | | |
|---|---|---|---|---|---|---|
|  |  | 50% Narrow | 90% Narrow | 90% Broad | 99% Narrow | 99% Broad |
| Median | 10% | 5 | 10 | 15 | 25 | 30 |
|  | 50% | 10 | 25 | 32.5 | 50 | 50 |
|  | 90% | 28.5 | 50 | 60 | 99 | 100 |
| Mean | 10% | 8.567 | 33.65 | 27.88 | 41.79 | 57.75 |
|  | 50% | 21.99 | 65.24 | 60.58 | 78.10 | 106.2 |
|  | 90% | 75.44 | 184.2 | 146.4 | 174.8 | 218.7 |
| Median Aggregate Forecast | 10% | 3.357 | 10.15 | 16.07 | 22.09 | 28.31 |
|  | 50% | 11.63 | 24.76 | 33.16 | 50.42 | 56.46 |
|  | 90% | 28.36 | 49.42 | 59.54 | 96.63 | 99.13 |

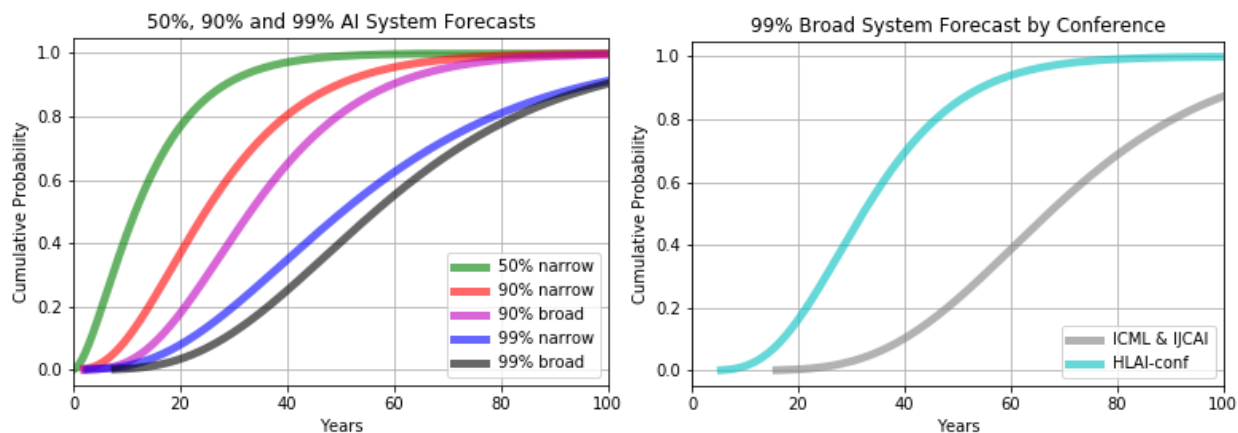

**Figure 2:** left – a) a plot comparing the five aggregate forecasts for each of the transformative AI scenarios of interest; right – b) a plot comparing the aggregate forecasts for the HLAI conference and the ICML and IJCAI conferences

Of the 164 completed responses, 15 participants gave at least one response that, using the criterion cited earlier (*i.e.*, forecasts of more than 5,000 years), was taken as indication that they believed this scenario would never happen. For both of the 99% scenarios all 15 participants indicated never to be a possibility, with 8, 6 and 5 indicating it to be a possibility for 90% broadly capable, 90% narrow and 50% narrow systems, respectively. In the comment box, some participants indicated that their extreme responses were intended to mean that this scenario would never happen or was not possible. For at least one scenario, 7 more of the 164 completed surveys exhibited inconsistencies (as discussed previously) that precluded further analysis. However, all completed expert surveys were considered for each individual question, and data was only disregarded if it deemed unsuitable for analysis of that question.

Figures 3a and 3b depict comparisons of all narrow system forecasts and both broadly capable system forecasts, respectively. These figures include a colored in region indicating a 95% confidence interval. The 95% confidence intervals were computed for each scenario using bootstrapping (Efron and Tibshirani 1994), with *n* equal to 10,000. *These plots illustrate experts' perceived timelines for increasing levels of transformative societal change due to AI progress over the next century.*

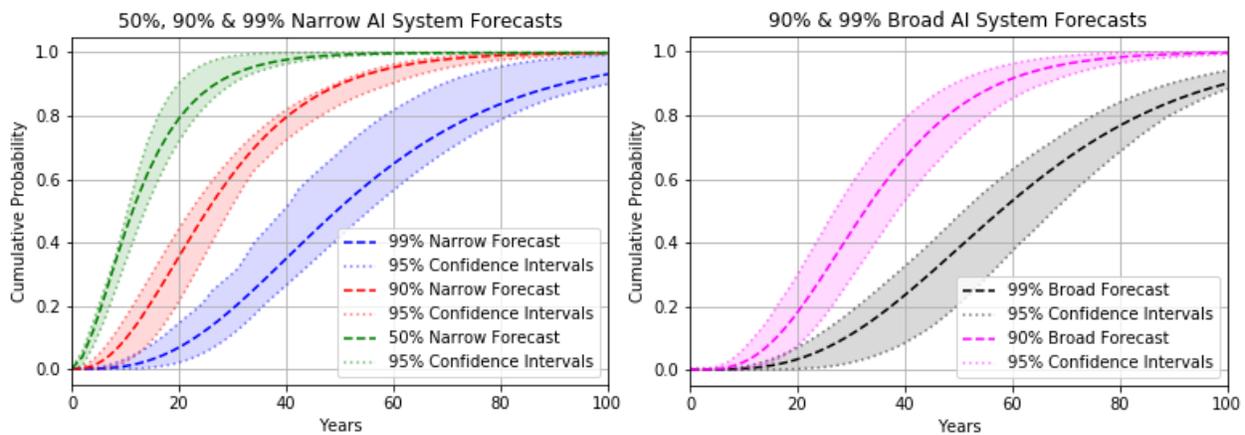

**Figure 3:** left – a) A plot of the aggregate forecasts for all narrow AI systems that includes 95% confidence intervals; right – b) A plot of the aggregate forecasts for both broad AI systems that includes 95% confidence intervals

**Statistical Analysis**

A multivariate regression model was used to test the impact of occupational role, region of residency, conference of attendance and gender. Six occupational roles were selected because they were indicative of different types of AI expertise and experience; these roles were graduate student, academic, researcher, engineer, executive and founder. 7 of the 164 respondents who completed the survey did not fall into any of the categories. Their roles included human resources, sales, journalist, etc. These seven respondents were considered to be non-experts and their data was discarded. Rather than looking at respondents' country of origin (Grace, Salvatier et al. 2018), we considered respondents' current region of residency (*i.e.*, continent). For this, Africa, South America, Oceania and respondents of undetermined regions were all grouped into a single category.

Due to the presence of multiple independent variables and covariance, the seemingly unrelated regressions (SUR) model (Zellner 1962) was used. SUR is a more efficient version of ordinary least square (OLS) for multiple equation systems; the potential correlation among the errors ($\varepsilon_{median}$ and $\varepsilon_{uncertainty}$) can be accounted for by using the SUR. When the error terms I the set of equations are uncorrelated, the model reduces to a set of OLS regression estimates (Green 2000).

We selected the log of the median forecast, denoted $Y_m$, as the first dependent variable. Then, rather than using the 10% and 90% forecasts as separate dependent variables, we combined them into a measure of uncertainty, $Y_u$, by subtracting the 10% forecast from the 90% forecast and taking the log. These values and the regression model can be seen in equations 1-4, where $F$ is the raw forecast value input by the respondent. The five regression models were fit using the R package systemfit for estimating systems of simultaneous equations (Henningsen and Hamann 2007). The results can be seen in Table II. Results for $Y_m$ are shown in the upper half of the table while results for $Y_u$ are shown in the bottom half of the table. The results that were found to be statistically significant are highlighted with shades of grey to indicate the level of significance (a key is located at the bottom of the table). The *p*-values that were found to be statistically significant for both dependent variables are italicized.

$$Y_m = \log(F_{50\%}) \tag{1}$$

$$Y_u = \log(F_{90\%} - F_{10\%}) \tag{2}$$

$$Y_m = \beta_0 + \beta_{1:5} X_{role} + \beta_{6:8} X_{region} + \beta_9 X_{conference} + \beta_{10} X_{gender} + \varepsilon_{median} \tag{3}$$

$$Y_u = \beta_0 + \beta'_{1:5} X_{role} + \beta'_{6:8} X_{region} + \beta'_9 X_{conference} + \beta'_{10} X_{gender} + \varepsilon_{uncertainty} \tag{4}$$

**Table II: Results from the Seemingly Unrelated Regressions Model**

|  |  | System Type | | | | | | | | | |
|---|---|---|---|---|---|---|---|---|---|---|---|
|  |  | 50% narrow | | 90% narrow | | 90% broad | | 99% narrow | | 99% broad | |
|  |  | β | p | β | p | β | p | β | p | β | p |
| $Y_m$ | Student | 0.697 | 0.093 | 0.266 | 0.542 | 0.503 | 0.190 | 0.130 | 0.755 | 0.082 | 0.851 |
|  | Academic | 0.480 | 0.248 | 0.023 | 0.958 | 0.312 | 0.417 | -0.050 | 0.904 | -0.087 | 0.841 |
|  | Research | 0.729 | *0.070* | 0.232 | 0.580 | 0.382 | 0.301 | -0.085 | 0.831 | -0.095 | 0.820 |
|  | Engineer | 0.606 | 0.153 | 0.435 | 0.329 | 0.517 | 0.187 | -0.152 | 0.724 | -0.107 | 0.812 |
|  | Executive | 0.160 | 0.732 | 0.518 | 0.296 | 0.538 | 0.216 | -0.035 | 0.942 | 0.228 | 0.642 |
|  | Europe | -0.041 | 0.887 | 0.070 | 0.817 | 0.157 | 0.566 | -0.594 | 0.042 | -0.498 | 0.102 |
|  | NA | -0.177 | 0.578 | -0.270 | 0.426 | -0.096 | 0.753 | -0.674 | 0.041 | -0.668 | 0.052 |
|  | Aisa | 0.240 | 0.488 | 0.342 | 0.356 | 0.523 | 0.108 | -0.259 | 0.456 | -0.154 | 0.670 |
|  | HLAI | -0.378 | *0.061* | -0.676 | *0.002* | -0.687 | *0.000* | -0.624 | *0.002* | -0.774 | *0.000* |
|  | Female | -0.392 | 0.166 | -0.061 | 0.843 | -0.490 | *0.079* | -0.259 | 0.379 | -0.367 | 0.233 |
| $Y_u$ | Student | 0.877 | 0.101 | 0.129 | 0.808 | 0.320 | 0.518 | 0.227 | 0.677 | 0.317 | 0.570 |
|  | Academic | 0.610 | 0.254 | -0.014 | 0.979 | 0.071 | 0.887 | -0.043 | 0.937 | 0.044 | 0.937 |
|  | Research | 1.003 | *0.053* | 0.320 | 0.532 | 0.318 | 0.506 | 0.381 | 0.467 | 0.438 | 0.414 |
|  | Engineer | 0.969 | 0.077 | 0.309 | 0.570 | 0.411 | 0.416 | -0.227 | 0.687 | -0.044 | 0.939 |
|  | Executive | 0.602 | 0.318 | 0.173 | 0.775 | 0.297 | 0.596 | 0.152 | 0.809 | 0.437 | 0.488 |
|  | Europe | 0.067 | 0.857 | 0.090 | 0.808 | 0.196 | 0.579 | -0.414 | 0.277 | -0.485 | 0.214 |
|  | NA | -0.063 | 0.877 | 0.053 | 0.899 | 0.062 | 0.876 | -0.375 | 0.380 | -0.528 | 0.229 |
|  | Aisa | 0.755 | 0.092 | 0.927 | 0.042 | 0.966 | 0.023 | 0.128 | 0.778 | 0.051 | 0.913 |
|  | HLAI | -0.527 | *0.043* | -0.855 | *0.001* | -0.705 | *0.004* | -0.624 | *0.019* | -0.758 | *0.005* |
|  | Female | -0.324 | 0.372 | -0.323 | 0.389 | -0.618 | *0.087* | -0.268 | 0.485 | -0.559 | 0.157 |
|  | Significance | 0 | | 0.1 | | 0.05 | | 0.01 | | 0.005 | |

## Naïve Calibration

Not all of the calibration questions can be evaluated at this time, however, an alternate technique for utilizing the calibration questions was employed. We call this very simple technique naïve calibration. It is founded on the assumption that any forecast with absolute certainty, i.e. 0% or 100%, is fundamentally flawed unless the forecaster has inside information. There is inherent uncertainty in any forecasts about the future, and the naïve calibration technique simply separates individuals who gave forecasts of absolute certainty in any one of the calibration questions from the remainder of the sample. A total of 41 of the 157 experts who completed the survey made at least one forecast of absolute certainty in the calibration questions. The SUR model was run again using the naïve calibration as a dummy variable. The results are shown in Table III.

**Table III: Results from the Seemingly Unrelated Regressions Model w/ Naïve Calibration**

| | | 50% narrow | | 90% narrow | | 90% broad | | 99% narrow | | 99% broad | |
|---|---|---|---|---|---|---|---|---|---|---|---|
| | | $\beta$ | $p$ | $\beta$ | $p$ | $\beta$ | $p$ | $\beta$ | $p$ | $\beta$ | $p$ |
| $Y_m$ | Student | 0.746 | 0.072 | 0.292 | 0.507 | 0.539 | 0.164 | 0.176 | 0.673 | 0.134 | 0.759 |
| | Academic | 0.513 | 0.214 | 0.040 | 0.928 | 0.320 | 0.408 | -0.051 | 0.903 | -0.098 | 0.823 |
| | Research | 0.817 | 0.042 | 0.277 | 0.513 | 0.429 | 0.251 | -0.031 | 0.939 | -0.043 | 0.918 |
| | Engineer | 0.619 | 0.141 | 0.439 | 0.326 | 0.527 | 0.180 | -0.142 | 0.741 | -0.097 | 0.829 |
| | Executive | 0.184 | 0.695 | 0.532 | 0.287 | 0.567 | 0.196 | 0.022 | 0.963 | 0.292 | 0.555 |
| | Europe | 0.007 | 0.982 | 0.094 | 0.760 | 0.195 | 0.483 | -0.541 | 0.067 | -0.436 | 0.156 |
| | NA | -0.156 | 0.627 | -0.258 | 0.451 | -0.072 | 0.816 | -0.639 | 0.053 | -0.623 | 0.071 |
| | Aisa | 0.254 | 0.461 | 0.348 | 0.349 | 0.535 | 0.102 | -0.247 | 0.477 | -0.141 | 0.697 |
| | HLAI | -0.317 | 0.116 | -0.645 | 0.003 | -0.654 | 0.001 | -0.586 | 0.005 | -0.737 | 0.001 |
| | Female | -0.397 | 0.164 | -0.066 | 0.834 | -0.516 | 0.070 | -0.305 | 0.308 | -0.430 | 0.169 |
| | Calibration | -0.026 | 0.947 | -0.001 | 0.995 | 0.067 | 0.680 | 0.143 | 0.434 | 0.212 | 0.265 |
| $Y_u$ | Student | 0.967 | 0.072 | 0.204 | 0.702 | 0.423 | 0.390 | 0.347 | 0.517 | 0.455 | 0.403 |
| | Academic | 0.632 | 0.237 | -0.009 | 0.987 | 0.054 | 0.912 | -0.074 | 0.890 | 0.006 | 0.991 |
| | Research | 1.110 | 0.033 | 0.401 | 0.437 | 0.416 | 0.381 | 0.500 | 0.332 | 0.567 | 0.280 |
| | Engineer | 0.991 | 0.070 | 0.333 | 0.540 | 0.452 | 0.366 | -0.206 | 0.708 | -0.019 | 0.972 |
| | Executive | 0.695 | 0.253 | 0.242 | 0.690 | 0.408 | 0.463 | 0.332 | 0.591 | 0.614 | 0.319 |
| | Europe | 0.154 | 0.682 | 0.166 | 0.658 | 0.318 | 0.369 | -0.265 | 0.481 | -0.318 | 0.404 |
| | NA | 0.017 | 0.967 | 0.106 | 0.799 | 0.152 | 0.699 | -0.269 | 0.521 | -0.406 | 0.343 |
| | Aisa | 0.787 | 0.079 | 0.935 | 0.040 | 1.008 | 0.016 | 0.156 | 0.725 | 0.085 | 0.850 |
| | HLAI | -0.470 | 0.073 | -0.797 | 0.003 | -0.636 | 0.009 | -0.542 | 0.039 | -0.665 | 0.013 |
| | Female | -0.389 | 0.292 | -0.387 | 0.310 | -0.739 | 0.042 | -0.425 | 0.267 | -0.739 | 0.058 |
| | Calibration | 0.413 | 0.421 | 0.186 | 0.415 | 0.371 | 0.074 | 0.508 | 0.031 | 0.609 | 0.011 |
| | Significance | 0 | | 0.1 | | 0.05 | | 0.010 | | 0.0050 | |

The results depicted in Table III were obtained through the same analysis as those in Table II, only with the inclusion of an 11[th] independent variable: the dummy variable for the naïve calibration. Table III is organized identically for ease of comparison. Table IV depicts the $R^2$ values for each of the dependent variables for each of the regressions in the SUR. The table is arranged to enable comparison of the $R^2$ values with and without the Naïve calibration.

**Table IV: Comparison of $R^2$ Values for the Original SUR and the Naïve Calibration SUR**

|  |  | $R^2$ |  |
|---|---|---|---|
|  |  | Original SUR | Naïve Calibration SUR |
| 99% Broadly Capable System | $Y_m$ | 0.0751 | 0.0936 |
|  | $Y_u$ | 0.0561 | 0.1143 |
| 99% Narrow Systems | $Y_m$ | 0.0751 | 0.0621 |
|  | $Y_u$ | 0.0561 | 0.0812 |
| 90% Broadly Capable System | $Y_m$ | 0.1632 | 0.1519 |
|  | $Y_u$ | 0.1307 | 0.1413 |
| 90% Narrow Systems | $Y_m$ | 0.0782 | 0.0651 |
|  | $Y_u$ | 0.1073 | 0.0996 |
| 50% Narrow Systems | $Y_m$ | 0.0612 | 0.0574 |
|  | $Y_u$ | 0.0961 | 0.0839 |

## Computational Resources

The final questions concerned the impact of computational resources on AI progress. The results for these questions can be seen in Figure 4a and Figure 4b. Figure 4a shows the survival function for the continuation of the exponential increase identified between AI milestones and computational training costs (Amodei and Hernandez 2018). Figure 4b shows the results, on a Likert-type scale, of participants responses regarding the hypothetical rate of increase in AI progress if all researchers were given access to unlimited computational resources.

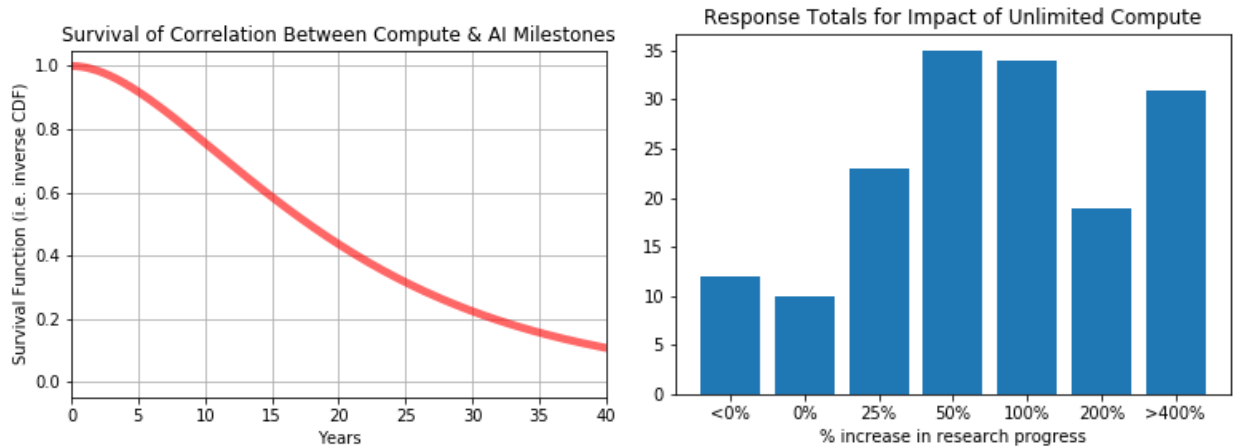

**Figure 4**: left – a) the survival function for the probability of continuing exponential increases in computational resources associated with future AI milestones; right – b) response totals for how much AI research will increase if everyone was given access to unlimited computational resources.

## Discussion

### Fixed Years Human Tasks

The results for the fixed years human tasks questions, depicted in Figures 1a and 1b, indicate that experts believe a large number of human tasks are already automatable, and that this number will grow substantially over the next five to ten years. Specifically, the median percentage of human tasks believed to be automatable now is 21.5%, and this number increases to 40% and 60% in five and ten years, respectively. However, the distributions for responses to this question cover the entire range of possibilities, which indicates significant variance and minimal value for quantitative forecasting. Despite this, it is noteworthy to observe in Figure 1a that the first two distributions are positive skewed while the last distribution is negative skewed. This shift suggests some sort of consensus toward transformative progress in automation of human tasks over the coming decade.

**Forecasts**

The survey reported here offers a reframing of the problem of forecasting AI progress with respect to previous studies by developing forecasts for five unique transformative AI scenarios. Some of these scenarios were intended to be similar to CAIS while others were intended to more closely approximate notions of AGI. We see this as addressing a limitation of previous studies. Significant AI progress and societal transformation could affect future progress profoundly through complex mechanisms (Gruetzemacher 2018). Thus, forecasts for more imminent and comprehensible scenarios of transformative AI progress are not only prudent but necessary. It is possible that AI experts who work on narrow systems can forecast requisite technologies for more extreme transformative scenarios without envisioning recursive research and development models, state funded efforts on the scale of the Manhattan Project or discontinuous change in the cost of computation. The implications of this reframing are plainly illustrated in Figures 3a and 3b, which compare the narrow and broad forecasts for the different transformative AI scenarios explored here. Thinking of these scenarios as levels of transformative AI, we see that the lower levels of narrow transformative AI have the earliest timelines and the least uncertainty.

**Statistical Analysis**

We incorporated all dependent variables into a multivariate model by combining the median forecast and the uncertainty. We conducted a Breusch-Pagan test (Breusch and Pagan 1979) for each of the scenarios, and one indicated the presence of heteroscedasticity. However, it is important to note that when the Breusch-Pagan test fails, the SUR model simply becomes separated OLS models because the separate modeling approach is a special case of the multivariate modeling approach. Thus, it is suitable to use SUR in all cases.

It can be seen in Table II that all of the *p*-values reported for HLAI are italicized indicating that, for each scenario, the impact of both the median and the uncertainty on the model was found to be statistically significant. These results show particularly strong statistical significance for the HLAI forecasts of scenarios for all systems of 90% and above. It can also be seen that the *β* coefficients for HLAI forecasts are negative for both the median and uncertainty over all scenarios. This indicates that HLAI participants expect all scenarios to happen sooner than participants of the other conferences. Furthermore, it indicates that HLAI participants have less uncertainty in their forecasts than participants of the other conferences. The forecasts for a 99% broadly capable system for participants of HLAI and ICML/IJCAI are depicted for comparison in Figure 2b.

Other paired correlations of interest (also italicized in Table II) include researcher's 50% narrow systems forecasts and females 90% narrow systems forecasts. The statistical significance is weak, but, limiting factors like female sample size (7) or the challenges of coding the roles of AI experts may obscure other underlying correlations. Also of interest are unpaired correlations for region of residence as well as for occupational role in forecasts of 50% narrow systems.

**Naïve Calibration**

The inclusion of the dummy variable created for the naïve calibration technique slightly reduced the *p*-values for the HLAI participants, however, the *β* coefficients are still negative for both the mean and the uncertainty over all scenarios. Despite this it is apparent by considering Table IV that the naïve calibration did have a significant effect on the model. The most significant effects can be seen for the case of the 99% broadly capable system forecasts, where the $R^2$ value increases for both the median and uncertainty. This indicates that the inclusion of the naïve calibration dummy variable increases the model's explanation of the variance by 1.85% and 5.82% for the mean and uncertainty, respectively. However, for the other cases the inclusion of the naïve calibration criterion is not as easily interpreted. This could easily be explained by the different types of forecasts being made – the 99% broadly capable system forecast is the one expected to be most susceptible to forecast quality – but there are many other possible explanations as well. These results are ultimately inconclusive, and we suggest that future work assess the viability of this naïve calibration technique in a more carefully designed experiment.

**Computational Resources**

Of final interest in our discussion are the results from the computational resource questions. The median aggregate forecast computed using the survival function indicates a greater than 75% probability of the trend identified by Amodei and Hernandez (Amodei and Hernandez 2018) continuing for ten years. Regarding the possibility of unlimited computational resources there appeared to be a general consensus of a positive impact on the rate of progress in AI research, with 51.2% of the participants indicating a rate of increase of 100% or greater. The consensus among researchers regarding the impact of computational resources on AI research is noteworthy and suggests that efforts should be made to explore expert opinion of these dynamics in future work. Considered with the results depicted in Figure 1, there appears to be some consensus that AI experts believe the next decade will continue to see AI progress proceed at a pace consistent with that of the past six years.

## Conclusions

We have presented a set of forecasts for five scenarios of transformative AI. In sum, we conclude that continuing advances in AI technology should be treated as something that will have transformative societal impact by the end of the decade. By design, this study does not lead to any conclusions regarding the economic impact of this possible transformation. Rather, it is intended to bring attention and awareness to the public as well as policy makers that a multitude of scenarios exists, some of which may not be far off and may present unforeseen risks and consequences. Future work should examine such scenarios more closely.

## Acknowledgements

We would like to thank Baobao Zhang, Adam Gleave, Danny Hernandez, Joel Lehman, Seth Baum, Ryan Carey, Matt Fay, Rick Schwall and Miles Brundage for their comments and suggestions during discussions at different phases of the study.

## References

Amodei, D. and D. Hernandez (2018). AI and Compute. Blog post, OpenAI.

Baum, S. D., et al. (2011). "How long until human-level AI? Results from an expert assessment." Technological Forecasting and Social Change 78(1): 185-195.

Bostrom, N. (2014) Superintelligence. Oxford University Press.

Breusch, T. S. and A. R. Pagan (1979). "A simple test for heteroscedasticity and random coefficient variation." Econometrica: Journal of the Econometric Society: 1287-1294.

Clemen, R. T. and R. L. Winkler (1999). "Combining probability distributions from experts in risk analysis." Risk analysis 19(2): 187-203.

Drexler, K. E. (2019). Reframing Superintelligence. Oxford University, Oxford University.

Efron, B. and R. J. Tibshirani (1994). An introduction to the bootstrap, CRC press.

Grace, K. (2015). "AI Timeling Surveys." from https://aiimpacts.org/ai-timeline-surveys/.

Grace, K., et al. (2018). "When will AI exceed human performance? Evidence from AI experts." Journal of Artificial Intelligence Research(62): 729-754.

Green, W. H. (2000). Econometric Analysis (5th ed.), New York: Prentice Hall.

Gruetzemacher, R. (2018). "Rethinking AI Strategy and Policy as Entangled Super Wicked Problems."

Henningsen, A. and J. D. Hamann (2007). "systemfit: A package for estimating systems of simultaneous equations in R." Journal of Statistical Software 23(4): 1-40.

Hora, S. C., et al. (2013). "Median aggregation of distribution functions." Decision Analysis 10(4): 279-291.

Michie, D. (1973). "Machines and the theory of intelligence." Nature 241(23.02): 1973.

Müller, V. C. and N. Bostrom (2016). Future progress in artificial intelligence: A survey of expert opinion. Fundamental issues of artificial intelligence, Springer: 555-572.

Vincent, S. B. (1912). The functions of the vibrissae in the behavior of the white rat, University of Chicago.

Walsh, T. (2018). "Expert and non-expert opinion about technological unemployment." International Journal of Automation and Computing 15(5): 637-642.

Zellner, A. (1962). "An efficient method of estimating seemingly unrelated regressions and tests for aggregation bias." Journal of the American statistical Association **57**(298): 348-368.

Zhang, B. and Dafoe, A., 2019. Artificial intelligence: American attitudes and trends. *Available at SSRN 3312874*.

**Ross Gruetzemacher**
Ross a PhD Candidate at Auburn University. His primary research interest is forecasting AI progress. He has conducted a seminar on the topic at the Center for the Governance of AI at Oxford University and is currently visiting the Center for the Study of Existential Risk at Cambridge University for this work. He has published work on applied AI as well as AI forecasting in journals such as the Journal of the American Medical Informatics Society and Big Data and Cognitive Computing.

**Kang Bok Lee**
Kang Bok Lee is an assistant professor of Business Analytics. He holds a PhD in Business Administration with concentration on Business Analytics and minor in Marketing. His research has appeared in the Production and Operations Management, the Academy of Management Journal, the International Journal of Physical Distribution & Logistic Management, the International Journal of Research in Marketing, Public Administration Review, Decision Sciences, the Journal of Business Logistics, the European Journal of Marketing, Public Administration Review, and the Journal of Financial Economic and Policy.

**David Paradice**
David Paradice is Chair of the Department of Systems and Technology and Harbert Eminent Scholar in the Raymond J. Harbert College of Business. He has published over 50 articles and book chapters on the use of information systems in support of managerial problem formulation and has served on over 50 doctoral dissertation committees. Prior to joining Auburn University, Dr. Paradice was on the faculties at Florida State University and Texas A&M University. He has been a Department Chair, Senior Associate Dean, Center Director, and teaching award winner. He has also served on several corporate advisory boards and worked as a consultant.